\documentclass[
  reprint,
  amsmath,amssymb,
  aps,
  pra,
  superscriptaddress,
  longbibliography
]{revtex4-2}

\usepackage{graphicx}
\usepackage{dcolumn}
\usepackage{bm}
\usepackage[utf8]{inputenc}
\usepackage[T2A,T1]{fontenc}
\usepackage[english]{babel}
\usepackage[caption=false]{subfig}
\usepackage{soul}
\usepackage{xcolor}

\begin{document}

\title{Casimir effect in twisted photonic gratings with in-plane chirality}

\author{Natalia S. Salakhova}
\email{natalia.salakhova@skoltech.ru}
\affiliation{Skolkovo Institute of Science and Technology, Moscow 121205, Russia}

\author{Sergey A. Dyakov}
\email{s.dyakov@skoltech.ru}
\affiliation{Skolkovo Institute of Science and Technology, Moscow 121205, Russia}

\author{Ilia M. Fradkin}
\affiliation{Skolkovo Institute of Science and Technology, Moscow 121205, Russia}
\affiliation{Moscow Institute of Physics and Technology, Moscow Region 141701, Russia}

\author{Nikolay A. Gippius}
\affiliation{Skolkovo Institute of Science and Technology, Moscow 121205, Russia}

\date{\today}


\begin{abstract}
We investigate the Casimir effect in a system of two twisted photonic gratings made of uniaxially anisotropic materials. Two distinct configurations are explored: a stack of symmetric gratings and a stack of in-plane chiral grating, with the latter realized by choosing specific orientation of anisotropy axis relative to stripes. We apply the reflection-matrix-based Casimir–Lifshitz formalism to explore how twist angle, material anisotropy, and the separation between gratings influence Casimir energy, force and torque. Our calculations reveal that the equilibrium orientation of the gratings is governed by the anisotropy rotation angles, leading to a chiral configuration where the anisotropy axes of the upper and lower gratings are mutually parallel. These findings demonstrate that material anisotropy provides a powerful mechanism for controlling rotational alignment forces in nanophotonic systems.
\end{abstract}

\maketitle
The Casimir effect is a fundamental quantum phenomenon arising from electromagnetic field fluctuations in vacuum. It produces a measurable force between neutral objects separated by nanometer to micrometer distances. First predicted by Hendrik Casimir in 1948 for two perfectly conducting plates \cite{Casimir1948}, it was extended by Lifshitz to real materials with arbitrary dielectric properties \cite{lifshitz1956}. Later developments of the Casimir effect theory incorporated dissipative effects and anisotropy \cite{dzyaloshinskii1961, Barash1978en, barash1978moment, van1995casimir}. Theoretical frameworks such as the Fourier modal method in scattering-matrix form now allow accurate modeling of Casimir interactions in complex geometries, including gratings and layered nanostructures \cite{rahi2009scattering, Lambrecht2008, messina2011scattering} .

The behavior of the Casimir effect is dictated by the symmetry of the system. In symmetric configurations, such as a pair of parallel isotropic slabs, the Casimir interaction appears as attractive or repulsive normal force. When symmetry is broken, either by geometric patterning or by material anisotropy, the Casimir force can gain lateral components and generate rotational torque \cite{spreng2022recent}. The Casimir torque has been studied theoretically and experimentally \cite{munday2005torque, Banishev2013, antezza2020giant, somers2018measurement} in systems containing uniaxial anisotropic layers and one-dimensional corrugated slabs. Both material anisotropy \cite{munday2005torque} and geometric anisotropy \cite{antezza2020giant, Banishev2013} lead to the alignment of the layers' anisotropy axes either parallel or perpendicular to each other.
In the literature, the Casimir force and torque have also been extensively studied for various material types starting from stacks of birefringent plates \cite{munday2005torque,hu2024twist, somers2017conditions}, metamaterials~\cite{rosa2008casimirmeta}, two-dimensional (2D) materials \cite{rodriguez2022twisted}, piezoelectric \cite{le2024phonon}, magnetodielectrics~\cite{rosa2008casimir}, Weyl semimetals~\cite{chen2020chiral, farias2020casimir} and chiral media~\cite{butcher2012casimir, farias2020casimir}. A comprehensive review of material effects in Casimir and van der Waals interactions is provided at Ref.~\cite{woods2016materials}. The role of geometry of interacting objects has been investigated in many studies~\cite{broer2023interplay, rodriguez2024giant, rodriguez2011casimir, Lussange2012}.

One of the most interesting regimes in Casimir interaction is a repulsive Casimir force that can result in the stable relative positioning of two objects \cite{zhao2019stable,ge2024fabry} and quantum levitation \cite{zhao2011repulsive, dou2014casimir, kenneth2002repulsive}. 
While the Casimir interaction between most macroscopic objects is inherently attractive, the realization of a repulsive Casimir force requires very specific material and environmental conditions \cite{dzyaloshinskii1961,somers2017conditions}. An alternative and practically important approach exploits the fact that, at micro- and nanometer distances, the magnitude of the Casimir effect can become comparable to that of an electrostatic interaction. Recently, the interplay of Casimir and electrostatic forces has been studied theoretically \cite{krasnov2024analysis} and experimentally  \cite{hovskova2025casimir} for gold flakes. Self-stabilization of the system at a certain distance between the flakes \cite{hovskova2025casimir} and the Casimir-assisted self-alignment \cite{kuccukoz2024quantum,munkhbat2021tunable} has been demonstrated.


In this work, we consider the Casimir interaction between recently introduced one-dimensional chiral gratings \cite{dyakov2024chiral}. 
These gratings are composed of stripes of uniaxially anisotropic dielectric material in which the anisotropy axis is not parallel (nor perpendicular) to the stripes. This geometry lacks a vertical mirror symmetry plane, and thus is in-plane chiral. Having a system of two such gratings oriented at a certain angle (twist angle) to each other, we calculate the Casimir force and torque between them. We show that the Casimir interaction in this system enables the existence of an equilibrium state at a certain twist angle, which is independent of the distance between the gratings. In the studied case, this angle appears to be non-zero and corresponds to the parallel alignment of the anisotropy axes in the upper and lower gratings' materials. Furthermore, we show that the presence of an electrostatic interaction between the gratings results in an equilibrium configuration with respect to the distance and the twist angle, where all forces are balanced. Finally, we demonstrate that the equilibrium twist angle depends on the strength of the anisotropy of the material. 

In the following, we focus on gratings with a period $p = 400$ nm, stripes width $w = 0.5p$, and grating height $h = 200$ nm. Although previous studies \cite{Lussange2012,Lambrecht2008} have shown that geometric parameters influence the Casimir energy, in this work we intentionally hold them constant. This allows us to isolate and investigate the effects of material anisotropy and structural asymmetry on the Casimir interaction. As parameters of our problem, we consider the distance $g$ between the gratings, the angle between the anisotropy axis and the stripes $\theta$, and the twist angle $\alpha$, that is, the angle between the stripes directions in the upper and lower gratings Fig.~\ref{fig1}(a)-(b). In the main section, we consider the case when the anisotropy axis is rotated by an angle $\theta_1 = \theta$ in the upper grating and by $\theta_2 = -\theta$ in the lower grating. The results for case $\theta_1 = \theta_2$ are presented in the Supplementary Materials \cite{supp}.

\begin{figure}
    \centering
    \includegraphics[width=0.9\linewidth]{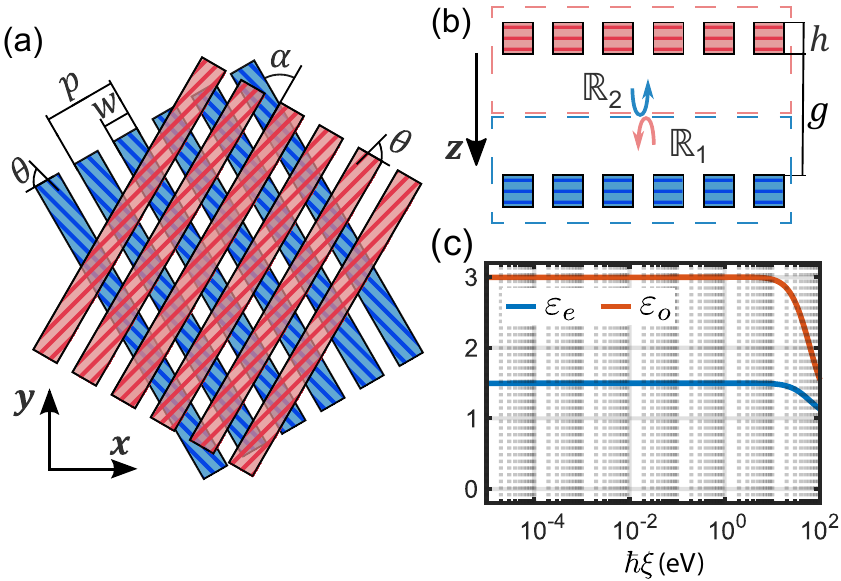}
    \caption{(a) Top and (b) side views of two twisted 1D photonic gratings separated by gap $g$, with anisotropy axes rotated by angles $+\theta$ and $-\theta$. (c) Dielectric permittivity components $\varepsilon_e$ and $\varepsilon_o$ at imaginary frequencies.}
    \label{fig1}
\end{figure}

As a material for gratings, we consider an artificial birefringent material with in-plane anisotropy and the dielectric permittivity tensor $\hat{\varepsilon}$ which in the principal axes has the form: 
\begin{equation}
    \hat{\varepsilon} = \begin{pmatrix}\varepsilon_e&0&0\\0&\varepsilon_o&0\\0&0&\varepsilon_o\end{pmatrix},~\varepsilon_{e(o)} = 1 + \frac{a_{e(o)}}{\hbar^2(\omega_{e(o)}^2 - \omega^2-i\omega\gamma_{e(o)})}, 
    \label{ten1}
\end{equation}
with the following parameters $a_e = 1.8$~eV$^2$,  $a_o = 7.2$~eV$^2$, $\hbar\omega_e = \hbar\omega_o = 60$~eV, $\hbar\gamma_e = \hbar\gamma_o = 5$~meV.  The chosen representation is a single oscillator form of the well-established multiple oscillator model, also known as the Ninham-Parsegian model ~\cite{mahanty1976dispersion,hough1980calculation,bergstrom1997hamaker}. This model satisfies the Kramers–Kronig relations and converges to unity in the high-frequency limit. More details can be found in the Supplementary Materials \cite{supp}.

We evaluate the Casimir energy per unit area using the scattering-matrix formalism. This approach allows one to accurately calculate the Casimir-Lifshitz interaction for bodies of arbitrary shape and material properties, provided that their reflection operators are known. Within the Casimir-Lifshitz scattering-matrix formalism \cite{Lambrecht2008, antezza2020giant}, the Casimir energy in the zero-temperature limit is given by:
\begin{equation}
    \mathcal{E} = \frac{\hbar}{8\pi^3}\int_{0}^{\infty }d\xi \\
\underset {\text{FBZ}}\int\ln \det \left[ \mathbb{I} - \mathbb{R}_1(i\xi, \mathbf{k}) \mathbb{R}_2(i\xi, \mathbf{k}) \right] \,d\mathbf{k}
\label{eq1}
\end{equation}
where $i\xi = \omega$ is the imaginary frequency, $\mathbf{k} = (k_x, k_y)$ is the in-plane wavevector integrated over the first Brillouin zone (see \cite{supp} Fig.S1 for details), and $\mathbb{R}_{1,2}$ are the reflection operators of the upper and lower gratings. Note that although the calculations are performed in the zero-temperature limit, the results are also applicable to objects at room temperature in case of small separation distances as in our study. The distance $g<200$~nm is much smaller than the thermal wavelength ($\lambda_T\approx7.63$ $\mu$m at room temperature) that makes quantum fluctuations the dominant contribution. The comparison of results obtained at $T = 0$~K and at room temperature presented in Fig.~S2 in Supplementary Materials \cite{supp}. 

In formula~\eqref{eq1}, reflection operators $\mathbb{R}_1$ and $\mathbb{R}_2$ are calculated using the Moire-adopted Fourier modal method developed for twisted one-dimensional gratings’ stacks \cite{salakhova2021fourier}. In calculations, we use $11$ Fourier harmonics for each grating or $121$ harmonics in total. 
In formula \eqref{eq1}, the exponential propagation factors, which are typically written explicitly~\cite{Lambrecht2008, antezza2020giant}, are already included in the definition of reflection operators, which are calculated in the center of the vacuum gap Fig.~\ref{fig1}(b). Thus, $\mathbb{R}_1$ and $\mathbb{R}_2$ depend not only on the twist angle $\alpha$, the anisotropy angle $\theta$ and the imaginary frequency $i\xi$, but also explicitly on the distance $g$. 

Knowing the Casimir energy $\mathcal{E}(\alpha,g)$ allows us to compute both the normal Casimir force and the Casimir torque. The force is obtained as a derivative of the energy with respect to the distance $g$, while the torque is given by the derivative with respect to the twist angle $\alpha$:
\begin{equation}
    \mathcal{F} = -\frac{\partial{\mathcal{E}}}{\partial{g}}, ~\mathcal{T} = -\frac{\partial{\mathcal{E}}}{\partial{\alpha}}
\end{equation}

\begin{figure}
    \centering
    \includegraphics[width=1\linewidth]{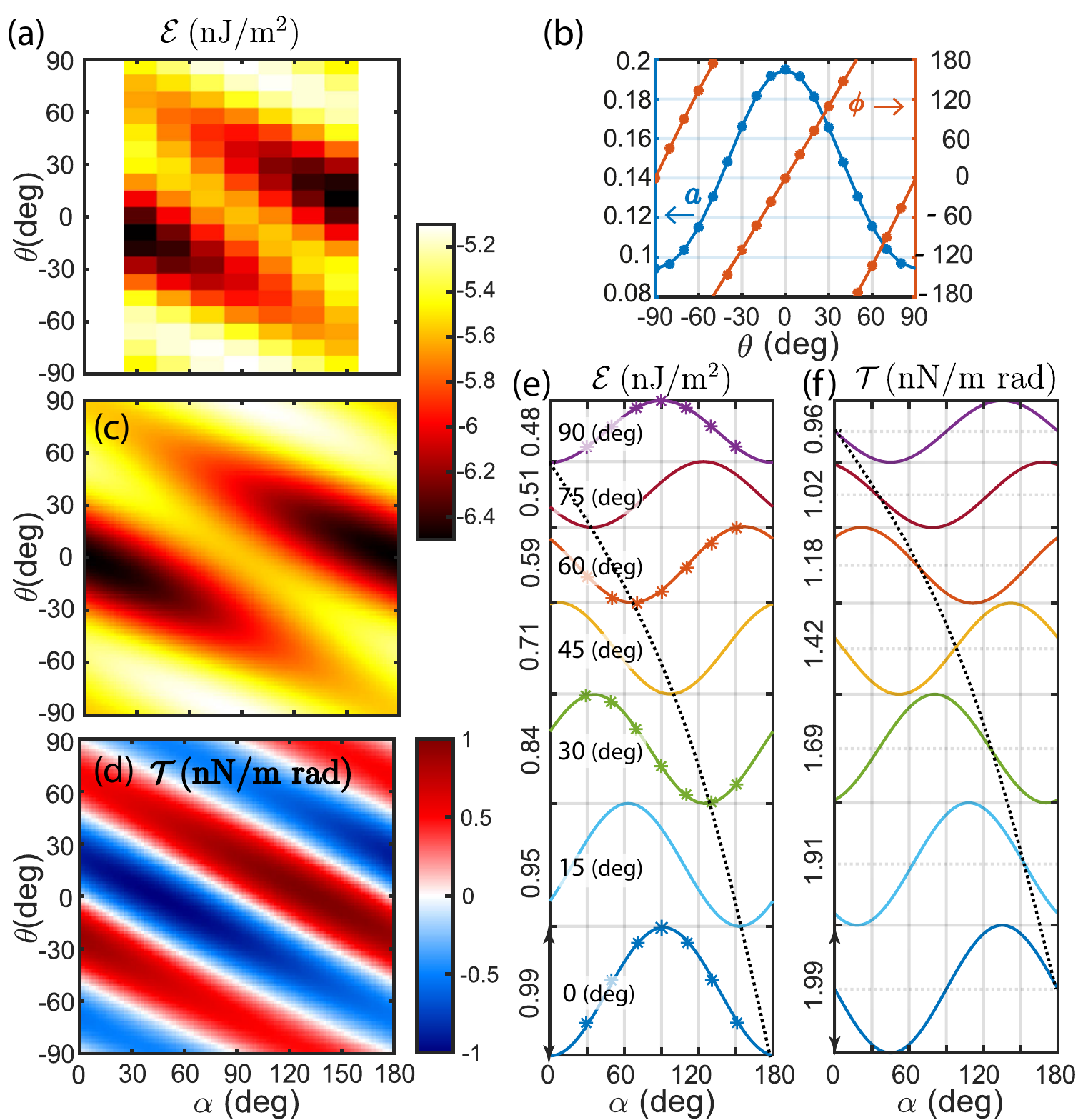}
    \caption{(a) Casimir energy versus twist angle $\alpha$ and anisotropy axis angles $\theta$ (raw data). (b) Fitting parameters $a$ and $\phi$ as a function of $\theta$.  (c)-(d) Energy and torque maps over $(\alpha,\theta)$ (analytical fits). (e)  Casimir energy $\mathcal{E}$ for different $\theta$. Stars denote numerical data, solid lines -  theoretical fits. (f) Casimir torque calculated from fitted curves. In calculations the gap thickness is $g = 100$~nm.}
    \label{fig2}
\end{figure}
Using this strategy, we compute the Casimir energy for a series of configurations defined by different twist angles $\alpha$ and anisotropy axis angles $\theta$ (see Fig. \ref{fig2}(a)).

We begin with the case $\theta = 0^\circ$, where both gratings have their anisotropy axis perpendicular to the gratings stripes. This configuration qualitatively corresponds to two twisted anisotropic slabs or 1D photonic gratings made of isotropic material. The numerical results for this case are well approximated by the analytical function:
 \begin{equation}
     \mathcal{E}(\alpha) = -a\cos{2\alpha}+c
     \label{fit10}
 \end{equation}
where $a$ and $c$ are fitting parameters. Consequently, the Casimir torque is given by:
\begin{equation}
\mathcal{T}(\alpha) = -\frac{\partial \mathcal{E}}{\partial \alpha} = -2a \sin2{\alpha},
\label{fit20}
\end{equation}
which agrees with a well-established behavior of the Casimir torque between anisotropic plates \cite{Barash1978en, van1995casimir}.

When the anisotropy axis is rotated around the $z$-axis by an angle $\theta \neq 0^{\circ}, 90^{\circ}$, the in-plane symmetry is broken and the grating becomes in-plane chiral. As a result of this broken symmetry, the minimum of the Casimir energy is shifted away from the symmetric value $\alpha = 0^\circ$ as shown in Fig.~\ref{fig2}(c)-(e). This shift indicates that the energetically most-favorable (i.e., equilibrium) configuration corresponds to the non-zero twist angle, and the system tends to spontaneously align at this offset orientation.

For an arbitrary $\theta$, the angular dependence of the Casimir energy can be described by an analytical function similar to \eqref{fit10} but with an additional phase shift: 
\begin{equation}
\mathcal{E}(\alpha,\theta) = -a(\theta) \cos\left(2\alpha + \phi(\theta)\right) + c(\theta),
\label{fit11}
\end{equation}
where $\phi$ denotes the shift in the energy minima. The fitting of the numerical results by formula \eqref{fit11} leads to $\phi \approx 4\theta$, $a \sim \cos^2{\theta}+\text{const}$ (see Fig.~\ref{fig2}(b)) and $c\sim-\cos^2{\theta}+\text{const}$. The retrieved  ($\alpha, \theta$)-dependence of the Casimir energy is shown in Fig.~\ref{fig2}(c).

As $\theta$ is varied from $0$ to $\pi/2$, the position of the energy minimum shifts continuously from $\pi$ to $0$, as shown in Fig.~\ref{fig2}(e) by the black dotted line. From Eq.~\ref{fit11}, with $\phi \approx 4\theta$, the condition for the energy minimum is expresses as $\cos(2\alpha+4\theta) = 1$, which gives an equilibrium twist angle $\alpha_\mathrm{eq} \approx \pi n-2\theta$, where $n \in \mathbf{Z}$. This means that the equilibrium state is achieved at a twist angle $\alpha$ such that the anisotropy axes in the upper and lower gratings' materials are approximately parallel to each other. 
This is particularly interesting in light of previous studies, which showed that homogeneous anisotropic layers always orient parallel or perpendicular to their anisotropy axes \cite{munday2005torque}, while gratings of isotropic materials always orient along the directions of stripes~\cite{antezza2020giant}.

Additionally, as shown in~\cite{antezza2020giant} for two infinite gratings, the twist-angle dependence of the Casimir energy exhibits a removable discontinuity in the zero rotation configuration due to the degeneracy between laterally shifted states. A similar discontinuity appears in our system at $\alpha = 0^{\circ}$. However, as demonstrated above, in the case of in-plane chiral gratings ($\theta \neq 0^{\circ}, 90^{\circ}$), the equilibrium twist angle $\alpha_\mathrm{eq} \neq 0$, and therefore the equilibrium state is well defined.

The Casimir torque, obtained from \eqref{fit11} is expressed as:
\begin{equation}
\mathcal{T}(\alpha,\theta) = -\frac{\partial\mathcal{E}}{\partial\alpha}=-2a(\theta) \sin(2\alpha + \phi(\theta)),
\label{eq:torque_fit}
\end{equation}

Note that in contrast to Casimir energy, the $\theta$-dependence of the Casimir torque is determined only by $a(\theta)$ and $\phi(\theta)$. The amplitude of the Casimir energy variation $a(\theta)$ and consequently the maximum Casimir torque $2a(\theta)$ decreases with increasing $\theta$ (see Fig.~\ref{fig2}(d-f).
Indeed, considering that in our system $\varepsilon_e < \varepsilon_o$, it follows that the dielectric contrast of the grating or the effective anisotropy is maximized when $\theta = 0^\circ$ and minimized when $\theta = 90^\circ$. The map of the Casimir torque at angles $\theta$ and $\alpha$ is presented in Fig.~\ref{fig2}(d). One can see that for each $\theta$, there are two values of $\alpha$ where the torque is equal to zero. These points are equilibrium states, one of which is stable and corresponds to the local minimum of the Casimir energy while another one is unstable and corresponds to the local maximum (Fig.~\ref{fig2}(e)-(f)).

\begin{figure*}[t!]
    \centering
    \includegraphics[width=1\textwidth]{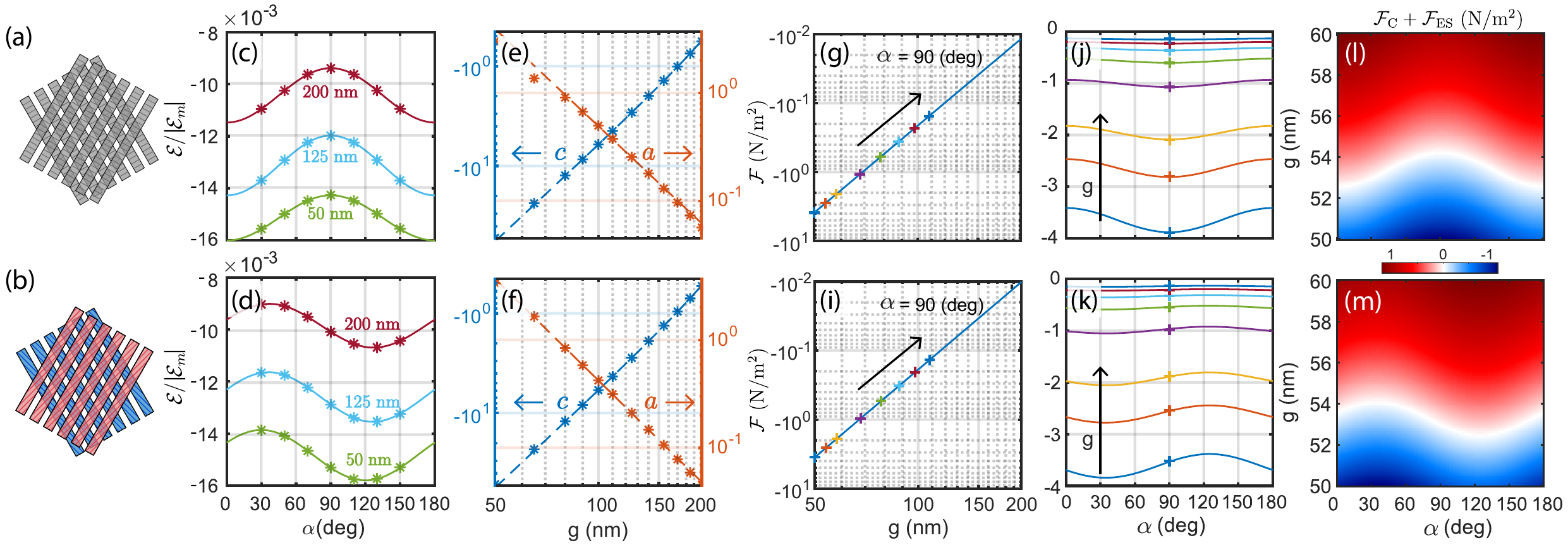}
    \caption{Stacks of (a) symmetric and (b) in-plane chiral gratings. (c)-(d): The twist-angle dependencies of the Casimir energy normalized to the absolute value of that for two ideal metallic plates for different gap sizes. (e)-(f): The gap-size dependencies of the fit parameters $a(g)$ and $c(g)$. The Casimir force versus (g)-(i) gap $g$ at $\alpha=90^\circ$ and (j)-(k) $\alpha$ for gaps marked in (g) and (i). (l)-(m) The resultant of the Casimir and electrostatic forces versus $g$ and $\alpha$. For in-plane chiral grating $\theta = 30^{\circ}$.}
    \label{fig3}
\end{figure*}
So far, we have calculated the Casimir energy at a fixed distance $g$. Now, let us study how the Casimir interaction strength scales with $g$ at a fixed angle of the anisotropy axis rotation $\theta$. For this, we calculate the $(\alpha,g)$-dependencies of the Casimir energy for the stack of symmetric lattices with $\theta = 0^{\circ}$ and for the stack of in-plane chiral lattices with $\theta = 30^{\circ}$ (see Fig.~\ref{fig3}). The Casimir energy is well described by an analytical expression:
\begin{equation}
\mathcal{E}(\alpha,g) = -a(g) \cos\left(2\alpha + \phi(g)\right) + c(g),
\label{eq:fit_g}
\end{equation}
where the fitting parameters $a(g)$, $\phi(g)$, and $c(g)$ are extracted numerically for every gap $g$.
 
The Casimir energy, normalized to that of two ideal metallic mirrors, $\mathcal{E}_m = -\frac{\pi^2 \hbar c}{720 g^3}$, is shown in Figs.~\ref{fig3}(c)--(d) as a function of the twist angle $\alpha$ for different $g$. The results show that the normalized energy changes with distance, exhibiting variation in both its amplitude and its average (offset) value. However, this variation is relatively small, which supports the assumption that, in our conditions, the Casimir energy decays with distance $g$ approximately according to the same power law as in the ideal metallic case. Note that while the Casimir energy itself depends on both the twist angle $\alpha$ and the gap $g$, the fitting coefficients $a$, $\phi$, and $c$ depend only on $g$. As seen in Fig.~\ref{fig3}(c)-(d), the position of energy minima and consecutively the phase shift $\phi(g)$ is constant $\phi(g) = \phi$, while the amplitude $a(g)$ and the offset $c(g)$ exhibit an approximate power-law dependence on $g$. These dependencies can be fitted by functions:
\begin{equation}
    a(g) = \frac{e_a}{g^{f_a}};~~c(g) = -\frac{e_c}{g^{f_c}}    
    \label{eq:ac}
\end{equation}
  where $e_a$,$e_c$, $f_a$, $f_c$ are the fitting parameters. The powers $f_a$ and $f_c$ are found to be close to 3: $f_a^\mathrm{symm} \approx 3$, $f_a^\mathrm{chir} \approx 3.1$, $f_c^\mathrm{symm} \approx 3.2, f_c^\mathrm{chir} \approx 3.2$) (see the solid lines in Fig.~\ref{fig3}(e), (f)).

Knowledge of the functions $a(g)$, $\phi$, and $c(g)$ enables us to reconstruct the dependence of the Casimir energy on the rotation angle $\alpha$ at any distance $g$, and to compute the Casimir force as its derivative with respect to the distance. Expression \eqref{eq:ac} gives:
\begin{align}
    \mathcal{E}(\alpha, g) &= -\frac{e_a}{g^{f_a}} \cos(2\alpha + \phi_0) - \frac{e_c}{g^{f_c}}, \\
    \mathcal{F}(\alpha, g) &= -\frac{\partial \mathcal{E}}{\partial g} = \frac{f_a e_a}{g^{f_a + 1}} \cos(2\alpha + \phi_0) + \frac{f_c e_c}{g^{f_c + 1}}.
    \label{eqfinalf}
\end{align}

Figure~\ref{fig3} (g)-(k) shows the Casimir force per unit area for the stack of symmetric and in-plane chiral gratings. Distance dependencies of the Casimir force Fig.~\ref{fig3} (g)-(i) follow the law expressed by formula \eqref{eqfinalf}. And important, the angular positions of the energy minimum remain unchanged for all values of $g$.  


At the nano- and micrometer scale, the Casimir force can be comparable to the electrostatic force \cite{munkhbat2021tunable,krasnov2024analysis,hovskova2025casimir}.
To investigate the interplay between these phenomena, we assume that the dielectric stripes of each grating carry a uniform surface charge density $\sigma$. The electrostatic force between the gratings was calculated using a simple model of Coulomb interaction between charges in the upper and lower gratings (see Supplementary Materials \cite{supp} for details). The force averaged over the superlattice unit cell was found to be independent on the twist angle $\alpha$, and varies only slowly within the considered range of distances $g$. 

Figure~\ref{fig3}(l)-(m) show the total interaction force, combining electrostatic repulsion and Casimir attraction, as a function of $\alpha$ and $g$ for stacks of symmetric and in-plane chiral gratings. 
One can see that for each twist angle $\alpha$ there is a distance $g_\mathrm{eq}$ where the total force equals zero. Although the electrostatic interaction is angle independent, the Casimir force exhibits a pronounced angular dependence; therefore, the equilibrium distance itself becomes angle dependent too, $g_\mathrm{eq} = g_\mathrm{eq}(\alpha)$. 


The stability of the equilibrium state with respect to the distance $g$ can be assessed from the $g$-dependencies of the forces. Since the attractive Casimir force decays as $\mathcal{F}_{\text{C}}\propto 1/g^4$ whereas the electrostatic repulsion $\mathcal{F}_\text{ES}$ is almost constant, the attractive contribution dominates at short distances, resulting in an unstable equilibrium state. This result is anticipated, given that the stable equilibrium would require the repulsive component to dominate at small distances and decrease faster than the attractive one, as demonstrated in the literature for metallic systems \cite{krasnov2024analysis, hovskova2025casimir}.
As for the twist angle stability, as discussed previously, in the symmetric case, the stable orientation corresponds to gratings parallel to each other, while in the chiral configuration, stability occurs at non-zero twist angles $\alpha_\mathrm{eq} = \pi n -2\theta$. Hence, in the latter case, the stack of in-plane chiral gratings can be stabilized in chiral configurations at distances $g_\mathrm{eq}(\alpha_\mathrm{eq})$.


\begin{figure}
    \centering
    \includegraphics[width=1\linewidth]{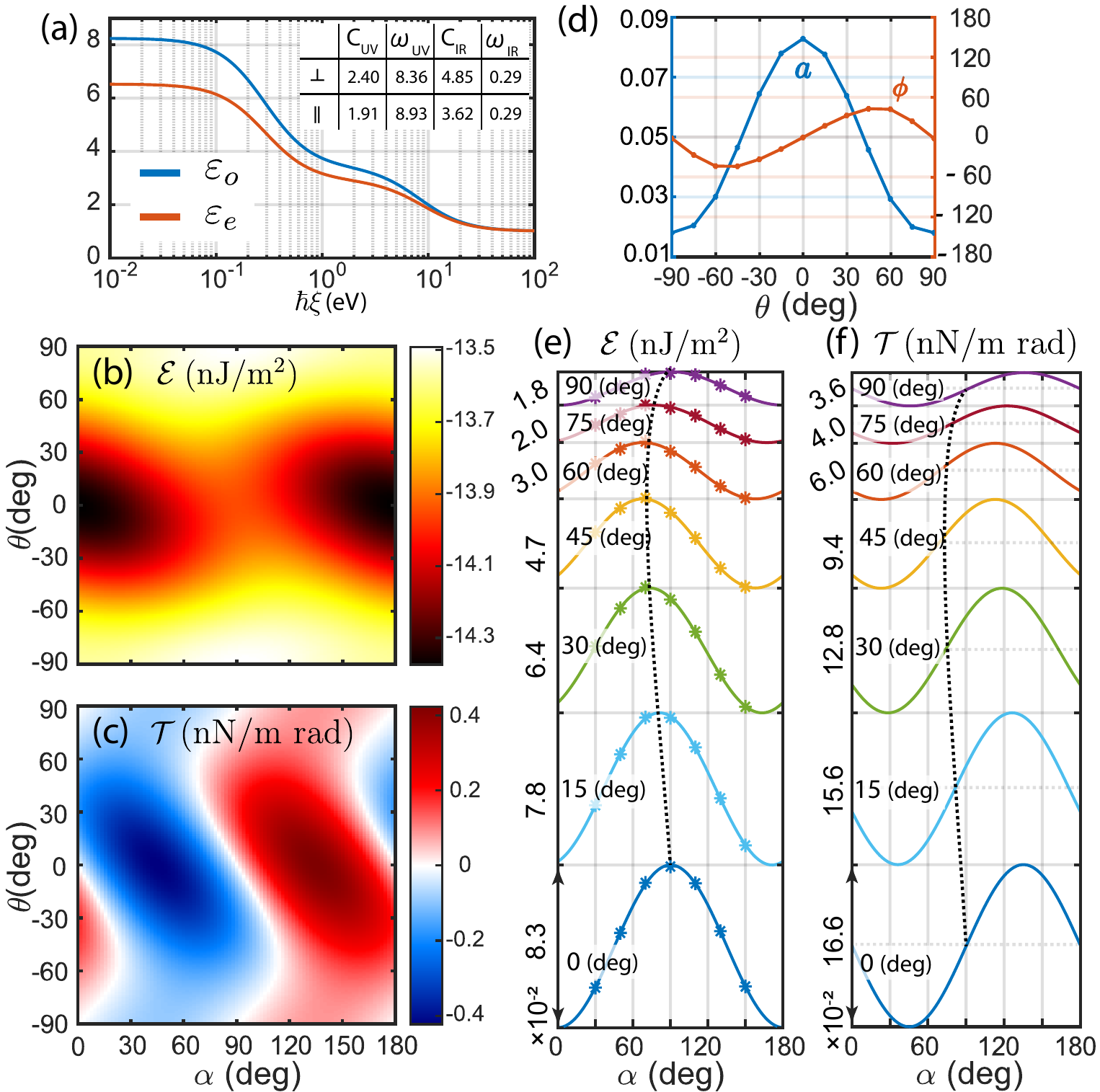}
    \caption{Casimir effect for grating made of LiIO$_3$. (a) Spectra of  dielectric permittivity tensor components; $\omega_{\text{UV}}$ and $\omega_{\text{IR}}$ in eV. (b) Casimir energy and (c) Casimir torque versus twist angle $\alpha$ and anisotropy axis angles $\theta$. (d) Fitting parameters $a$ and $\phi$ from Eq.~\ref{fit11} as a function of $\theta$. (e)  Casimir energy $\mathcal{E}$ and (f) Casimir torque for different $\theta$ (cross sections of panels (b) and (c)). Stars denote numerical data, solid lines --  theoretical fits. Results are obtained for distance between gratings $g = 100$~nm.}
    \label{figLiIO3}
\end{figure}

So far, we have considered the Casimir interaction between gratings composed of a model uniaxial anisotropic material with very high optical anisotropy. Let us now simulate the Casimir interaction using real birefringent material Refs.~\cite{mahanty1976dispersion,hough1980calculation,bergstrom1997hamaker}.
We choose a material that exhibits strong birefringence, with one of the permittivity components (ordinary or extraordinary) remaining larger than the other. One of the materials that satisfies these requirements is LiIO$_3$ with the dielectric tensor $\varepsilon = \text{diag}(\varepsilon_{\perp},~\varepsilon_{\parallel}, ~\varepsilon_{\parallel})$ where ordinary $\parallel$ and extraordinary $\perp$ components are described by the Ninham-Parsegian model. The parameters of the model were taken from Ref.~\cite{somers2017conditions}, and the corresponding permittivity spectra are shown in Fig.~\ref{figLiIO3}(a).

Figures~\ref{figLiIO3}(b,c) present the calculated Casimir energy and Casimir torque between gratings made of LiIO$_3$ fitted using Eqs.~\ref{fit11} and~\ref{eq:torque_fit}, with the fitting parameters plotted in Fig.~\ref{figLiIO3}(d). As shown in Fig.~\ref{figLiIO3}(e) and (f), the Casimir energy follows a cosine-like dependence on the relative rotation angle, while the Casimir torque exhibits a sine-like behavior as it was in the case of the artificial model material. The positions of the energy maxima and minima, as well as the zero-torque angles $\alpha_{\text{eq}}$, shift with the change of the tilt angle $\theta$.

Compared to the system with artificial model material, the phase parameter $\phi$ governing the angular shift for LiIO$_3$ is limited to values up to approximately $45^{\circ}$ (see Fig.~\ref{figLiIO3}(d)), rather than spanning the full range $[-\pi,\pi]$ (see Fig.~\ref{fig2}(b)). As a result, the maximum shift of the extrema is $\max{(\alpha_{\text{eq}})} = \phi/2 = 22.5^{\circ}$. Although the absolute Casimir energy is larger due to the higher permittivity of LiIO$_3$, its modulation parameter $a$ and consequently the Casimir torque is approximately one order of magnitude smaller.

These results demonstrate that the angular shift of the extrema points due to the interplay between material and geometry anisotropy is present both in engineered model materials (e.g. metamaterials) and in real anisotropic materials that exhibit broadband anisotropy. However, the severity of the effect depends on the strength of the material anisotropy.

In summary, we have investigated the Casimir interaction between twisted one-dimensional gratings exhibiting in-plane chirality induced by rotation of the anisotropy axes. Using the reflection-matrix-based Casimir–Lifshitz formalism, we calculated the Casimir energy as a function of the twist angle, material anisotropy, and separation distance between the gratings. Our results reveal the existence of a chiral equilibrium configuration with respect to the twist angle $\alpha$, which occurs when the anisotropy axes of the upper and lower gratings are nearly parallel. Furthermore, we demonstrated that a stable equilibrium with respect to both the twist angle and the separation distance can be achieved when the gratings carry an electric charge. Finally, we showed that angular stabilization of twisted grating stacks in a chiral configuration can also be realized in real anisotropic materials; however, the magnitude of the effect strongly depends on the strength of the material anisotropy. These findings show that twisted stacks of in-plane chiral photonic structures enable precise control of mechanical movements using Casimir forces, paving the way for nanoscale actuation, self-alignment, and reconfigurable chiral quantum and photonic systems.

\begin{acknowledgments}
This work was supported by the Russian Science Foundation (Grant No. 22-12-00351$\Pi$).

We thank Oleg Kotov for valuable recommendations and fruitful discussions regarding the history and the current state of the problem.
\end{acknowledgments}

Data availability \cite{data}.


\bibliography{Casimir_Ref}

@Article{Lussange2012,
  author    = {Lussange, J. and Guérout, R. and Lambrecht, A.},
  journal   = {Physical Review A},
  title     = {Casimir energy between nanostructured gratings of arbitrary periodic profile},
  year      = {2012},
  issn      = {1094-1622},
  month     = dec,
  number    = {6},
  pages     = {062502},
  volume    = {86},
  doi       = {10.1103/physreva.86.062502},
  publisher = {American Physical Society (APS)},
}

@article{dyakov2024chiral,
  title={Chiral light in twisted fabry--p{\'e}rot cavities},
  author={Dyakov, Sergey A and Salakhova, Natalia S and Ignatov, Alexey V and Fradkin, Ilia M and Panov, Vitaly P and Song, Jang-Kun and Gippius, Nikolay A},
  journal={Advanced Optical Materials},
  volume={12},
  number={12},
  pages={2302502},
  year={2024},
  publisher={Wiley Online Library}
}

@Article{Lambrecht2008,
  author    = {Lambrecht, Astrid and Marachevsky, Valery N.},
  journal   = {Physical Review Letters},
  title     = {Casimir Interaction of Dielectric Gratings},
  year      = {2008},
  issn      = {1079-7114},
  month     = oct,
  number    = {16},
  pages     = {160403},
  volume    = {101},
  doi       = {10.1103/physrevlett.101.160403},
  publisher = {American Physical Society (APS)},
}

@article{antezza2020giant,
  title={Giant Casimir torque between rotated gratings and the $\theta$= 0 anomaly},
  author={Antezza, Mauro and Chan, Ho Bun and Guizal, Brahim and Marachevsky, Valery N and Messina, Riccardo and Wang, Mingkang},
  journal={Physical review letters},
  volume={124},
  number={1},
  pages={013903},
  year={2020},
  publisher={APS}
}

@Article{Banishev2013,
  author    = {Banishev, A. A. and Wagner, J. and Emig, T. and Zandi, R. and Mohideen, U.},
  journal   = {Physical Review Letters},
  title     = {Demonstration of Angle-Dependent Casimir Force between Corrugations},
  year      = {2013},
  issn      = {1079-7114},
  month     = jun,
  number    = {25},
  pages     = {250403},
  volume    = {110},
  doi       = {10.1103/physrevlett.110.250403},
  publisher = {American Physical Society (APS)},
}

@Article{Casimir1948,
  author    = {Casimir, H. B. G. and Polder, D.},
  journal   = {Physical Review},
  title     = {The Influence of Retardation on the London-van der Waals Forces},
  year      = {1948},
  issn      = {0031-899X},
  month     = feb,
  number    = {4},
  pages     = {360--372},
  volume    = {73},
  doi       = {10.1103/physrev.73.360},
  publisher = {American Physical Society (APS)},
}

@article{lifshitz1956,
  author = {Lifshitz, E. M.},
  title = {The theory of molecular attractive forces between solids},
  journal = {Sov. Phys. JETP},
  volume = {2},
  pages = {73--83},
  year = {1956}
}

@article{dzyaloshinskii1961,
  author = {Dzyaloshinskii, I. E. and Lifshitz, E. M. and Pitaevskii, L. P.},
  title = {The general theory of van der Waals forces},
  journal = {Advances in Physics},
  volume = {10},
  number = {38},
  pages = {165--209},
  year = {1961},
  doi = {10.1080/00018736100101281}
}

@article{rahi2009scattering,
  author = {Rahi, Sahand Jamal and Emig, Thorsten and Graham, Noah and Jaffe, Robert L. and Kardar, Mehran},
  title = {Scattering theory approach to electrodynamic Casimir forces},
  journal = {Phys. Rev. D},
  volume = {80},
  pages = {085021},
  year = {2009},
  doi = {10.1103/PhysRevD.80.085021}
}

@article{messina2011scattering,
  title={Scattering-matrix approach to Casimir-Lifshitz force and heat transfer out of thermal equilibrium between arbitrary bodies},
  author={Messina, Riccardo and Antezza, Mauro},
  journal={Physical Review A—Atomic, Molecular, and Optical Physics},
  volume={84},
  number={4},
  pages={042102},
  year={2011},
  publisher={APS}
}

@article{munday2005torque,
  title={Torque on birefringent plates induced by quantum fluctuations},
  author={Munday, Jeremy N and Iannuzzi, Davide and Barash, Yuri and Capasso, Federico},
  journal={Physical Review A—Atomic, Molecular, and Optical Physics},
  volume={71},
  number={4},
  pages={042102},
  year={2005},
  publisher={APS}
}

@article{somers2018measurement,
  title={Measurement of the Casimir torque},
  author={Somers, David AT and Garrett, Joseph L and Palm, Kevin J and Munday, Jeremy N},
  journal={Nature},
  volume={564},
  number={7736},
  pages={386--389},
  year={2018},
  publisher={Nature Publishing Group UK London}
}

@article{chen2020chiral,
  title={Chiral-anomaly-driven Casimir-Lifshitz torque between Weyl semimetals},
  author={Chen, Liang and Chang, Kai},
  journal={Physical Review Letters},
  volume={125},
  number={4},
  pages={047402},
  year={2020},
  publisher={APS}
}

@article{butcher2012casimir,
  title={Casimir--Polder forces between chiral objects},
  author={Butcher, David T and Buhmann, Stefan Yoshi and Scheel, Stefan},
  journal={New Journal of Physics},
  volume={14},
  number={11},
  pages={113013},
  year={2012},
  publisher={IOP Publishing}
}

@article{van1995casimir,
  title={Casimir torque between dielectrics},
  author={Van Enk, SJ},
  journal={Physical Review A},
  volume={52},
  number={4},
  pages={2569},
  year={1995},
  publisher={APS}
}

@article{rodriguez2024giant,
  title={Giant anisotropy and Casimir phenomena: The case of carbon nanotube metasurfaces},
  author={Rodriguez-Lopez, Pablo and Le, Dai-Nam and Bondarev, Igor V and Antezza, Mauro and Woods, Lilia M},
  journal={Physical Review B},
  volume={109},
  number={3},
  pages={035422},
  year={2024},
  publisher={APS}
}

@article{broer2023interplay,
  title={Interplay between finite thickness and chirality effects on the Casimir-Lifshitz torque with nematic cholesteric liquid crystals},
  author={Broer, Wijnand and Podgornik, Rudolf},
  journal={Physical Review A},
  volume={108},
  number={1},
  pages={012814},
  year={2023},
  publisher={APS}
}

@article{Barash1978en,
  author       = {Barash, Yu. S.},
  title        = {Moment of van der Waals forces between anisotropic dielectrics},
  journal      = {Izvestiya Vysshikh Uchebnykh Zavedenii, Radiofizika},
  year         = {1978},
  volume       = {12},
  number       = {12},
  pages        = {1637--1644},
  note         = {in Russian},
  language     = {english}
}

@article{kenneth2002repulsive,
  title={Repulsive casimir forces},
  author={Kenneth, O and Klich, I and Mann, A and Revzen, M},
  journal={Physical review letters},
  volume={89},
  number={3},
  pages={033001},
  year={2002},
  publisher={APS}
}

@article{somers2017conditions,
  title={Conditions for repulsive Casimir forces between identical birefringent materials},
  author={Somers, David AT and Munday, Jeremy N},
  journal={Physical Review A},
  volume={95},
  number={2},
  pages={022509},
  year={2017},
  publisher={APS}
}

@article{hu2024twist,
  title={Twist-induced Casimir attractive-repulsive transition based on lithium iodate},
  author={Hu, Yang and Wu, Xiaohu and Liu, Haotuo and Ge, Wenxuan and Zhang, Jihong and Huang, Xiuquan},
  journal={ACS Photonics},
  volume={11},
  number={5},
  pages={1998--2006},
  year={2024},
  publisher={ACS Publications}
}

@article{salakhova2021fourier,
  title={Fourier modal method for moir{\'e} lattices},
  author={Salakhova, Natalia S and Fradkin, Ilia M and Dyakov, Sergey A and Gippius, Nikolay A},
  journal={Physical Review B},
  volume={104},
  number={8},
  pages={085424},
  year={2021},
  publisher={APS}
}

@article{krasnov2024analysis,
  title={Analysis of stability and near-equilibrium dynamics of self-assembled Casimir cavities},
  author={Krasnov, Mikhail and Mazitov, Arslan and Orekhov, Nikita and Baranov, Denis G},
  journal={Physical Review B},
  volume={109},
  number={19},
  pages={195411},
  year={2024},
  publisher={APS}
}

@article{hovskova2025casimir,
  title={Casimir self-assembly: A platform for measuring nanoscale surface interactions in liquids},
  author={Ho{\v{s}}kov{\'a}, Michaela and Kotov, Oleg V and K{\"u}{\c{c}}{\"u}k{\"o}z, Bet{\"u}l and Murphy, Catherine J and Shegai, Timur O},
  journal={Proceedings of the National Academy of Sciences},
  volume={122},
  number={31},
  pages={e2505144122},
  year={2025},
  publisher={National Academy of Sciences}
}

@article{munkhbat2021tunable,
  title={Tunable self-assembled Casimir microcavities and polaritons},
  author={Munkhbat, Battulga and Canales, Adriana and K{\"u}{\c{c}}{\"u}k{\"o}z, Bet{\"u}l and Baranov, Denis G and Shegai, Timur O},
  journal={Nature},
  volume={597},
  number={7875},
  pages={214--219},
  year={2021},
  publisher={Nature Publishing Group UK London}
}

@article{kuccukoz2024quantum,
  title={Quantum trapping and rotational self-alignment in triangular Casimir microcavities},
  author={K{\"u}{\c{c}}{\"u}k{\"o}z, Bet{\"u}l and Kotov, Oleg V and Canales, Adriana and Polyakov, Alexander Yu and Agrawal, Abhay V and Antosiewicz, Tomasz J and Shegai, Timur O},
  journal={Science Advances},
  volume={10},
  number={17},
  pages={eadn1825},
  year={2024},
  publisher={American Association for the Advancement of Science}
}

@article{rodriguez2011casimir,
  title={The Casimir effect in microstructured geometries},
  author={Rodriguez, Alejandro W and Capasso, Federico and Johnson, Steven G},
  journal={Nature photonics},
  volume={5},
  number={4},
  pages={211--221},
  year={2011},
  publisher={Nature Publishing Group UK London}
}

@article{farias2020casimir,
  title={Casimir force between Weyl semimetals in a chiral medium},
  author={Farias, M Bel{\'e}n and Zyuzin, Alexander A and Schmidt, Thomas L},
  journal={Physical Review B},
  volume={101},
  number={23},
  pages={235446},
  year={2020},
  publisher={APS}
}

@article{rosa2008casimirmeta,
  title={Casimir-Lifshitz theory and metamaterials},
  author={Rosa, FSS and Dalvit, DAR and Milonni, PW},
  journal={Physical review letters},
  volume={100},
  number={18},
  pages={183602},
  year={2008},
  publisher={APS}
}

@article{barash1978moment,
  title={Moment of van der Waals forces between anisotropic bodies},
  author={Barash, Yu S},
  journal={Radiophysics and Quantum Electronics},
  volume={21},
  number={11},
  pages={1138--1143},
  year={1978},
  publisher={Springer}
}

@article{rosa2008casimir,
  title={Casimir interactions for anisotropic magnetodielectric metamaterials},
  author={Rosa, FSS and Dalvit, DAR and Milonni, Peter W},
  journal={Physical Review A—Atomic, Molecular, and Optical Physics},
  volume={78},
  number={3},
  pages={032117},
  year={2008},
  publisher={APS}
}

@article{woods2016materials,
  title={Materials perspective on Casimir and van der Waals interactions},
  author={Woods, Lilia M and Dalvit, Diego Alejandro Roberto and Tkatchenko, Alexandre and Rodriguez-Lopez, Pablo and Rodriguez, Alejandro W and Podgornik, Rudolf},
  journal={Reviews of Modern Physics},
  volume={88},
  number={4},
  pages={045003},
  year={2016},
  publisher={APS}
}

@article{rodriguez2022twisted,
  title={Twisted bilayered graphenes at magic angles and Casimir interactions: correlation-driven effects},
  author={Rodriguez-Lopez, Pablo and Le, Dai-Nam and Calder{\'o}n, Mar{\'\i}a J and Bascones, Elena and Woods, Lilia M},
  journal={2D Materials},
  volume={10},
  number={1},
  pages={014006},
  year={2022},
  publisher={IOP Publishing}
}

@article{le2024phonon,
  title={Phonon-assisted Casimir interactions between piezoelectric materials},
  author={Le, Dai-Nam and Rodriguez-Lopez, Pablo and Woods, Lilia M},
  journal={Communications Materials},
  volume={5},
  number={1},
  pages={1--7},
  year={2024},
  publisher={Nature Publishing Group}
}

@book{mahanty1976dispersion,
  title={Dispersion Forces},
  author={Mahanty, J. and Ninham, B.W.},
  isbn={9780124650503},
  lccn={75027236},
  series={Colloid Science: a series of monographs},
  url={https://books.google.be/books?id=Fh2FAAAAIAAJ},
  year={1976},
  publisher={Academic Press}
}

@article{bergstrom1997hamaker,
  title={Hamaker constants of inorganic materials},
  author={Bergstr{\"o}m, Lennart},
  journal={Advances in colloid and interface science},
  volume={70},
  pages={125--169},
  year={1997},
  publisher={Elsevier}
}

@article{hough1980calculation,
  title={The calculation of Hamaker constants from Liftshitz theory with applications to wetting phenomena},
  author={Hough, David B and White, Lee R},
  journal={Advances in Colloid and Interface Science},
  volume={14},
  number={1},
  pages={3--41},
  year={1980},
  publisher={Elsevier}
}

@article{ge2024fabry,
  title={Fabry-P{\'e}rot nanocavities controlled by Casimir forces in electrolyte solutions},
  author={Ge, Lixin and Liu, Kaipeng and Gong, Ke and Podgornik, Rudolf},
  journal={Physical Review Applied},
  volume={21},
  number={4},
  pages={044040},
  year={2024},
  publisher={APS}
}

@article{guerout2013thermal,
  title={Thermal Casimir force between nanostructured surfaces},
  author={Gu{\'e}rout, Romain and Lussange, J and Chan, Ho Bun and Lambrecht, Astrid and Reynaud, Serge},
  journal={Physical Review A—Atomic, Molecular, and Optical Physics},
  volume={87},
  number={5},
  pages={052514},
  year={2013},
  publisher={APS}
}

@article{zhao2011repulsive,
  title={Repulsive Casimir forces with finite-thickness slabs},
  author={Zhao, R and Koschny, Th and Economou, EN and Soukoulis, CM},
  journal={Physical Review B—Condensed Matter and Materials Physics},
  volume={83},
  number={7},
  pages={075108},
  year={2011},
  publisher={APS}
}

@article{zhao2019stable,
  title={Stable Casimir equilibria and quantum trapping},
  author={Zhao, Rongkuo and Li, Lin and Yang, Sui and Bao, Wei and Xia, Yang and Ashby, Paul and Wang, Yuan and Zhang, Xiang},
  journal={Science},
  volume={364},
  number={6444},
  pages={984--987},
  year={2019},
  publisher={American Association for the Advancement of Science}
}

@article{dou2014casimir,
  title={Casimir quantum levitation tuned by means of material properties and geometries},
  author={Dou, Maofeng and Lou, Fei and Bostr{\"o}m, Mathias and Brevik, Iver and Persson, Clas},
  journal={Physical Review B},
  volume={89},
  number={20},
  pages={201407},
  year={2014},
  publisher={APS}
}

@article{spreng2022recent,
  title={Recent developments on the Casimir torque},
  author={Spreng, Benjamin and Gong, Tao and Munday, Jeremy N},
  journal={International Journal of Modern Physics A},
  volume={37},
  number={19},
  pages={2241011},
  year={2022},
  publisher={World Scientific}
}

@misc{supp,
    author       = {},
    title        = {Detailed descriptions of the material model, computational methodology, temperature dependence, and electrostatic force calculations are in the {S}upplementary {M}aterials},
    howpublished = {\url{https://doi.org/10.1103/6z4l-xb93}},
}

@misc{data,
    author       = {},
    title        = {All raw data corresponding to the findings reported in this manuscript are available upon reasonable request. Contact:natalia.salakhova@skoltech.ru, s.dyakov@skoltech.ru}
}

\title{Supplementary materials for "Casimir effect in twisted photonic gratings with in-plane chirality"}

\maketitle
\section{Material dielectric permittivity tensor}

To describe the dielectric permittivity of the material at all frequencies we use Eq.1 from the main text. The chosen representation is a single oscillator form of the well-established multiple oscillator model, also known as the Ninham--Parsegian model. This model is widely used for the description of the dielectric properties of various glasses and dielectrics, and although generally includes only two oscillators in the ultraviolet and infrared regions, it has been shown to provide good agreement between theory and experiment for many materials~\cite{mahanty1976dispersion,hough1980calculation,bergstrom1997hamaker}. Single-oscillator descriptions have also been successfully used for some real materials, for example, TiO$_2$ (see Ref.~\cite{bergstrom1997hamaker}). 

This model satisfies the Kramers–Kronig relations and converges to unity in the high-frequency limit.

\begin{figure*}
    \centering    \includegraphics[width=0.8\linewidth]{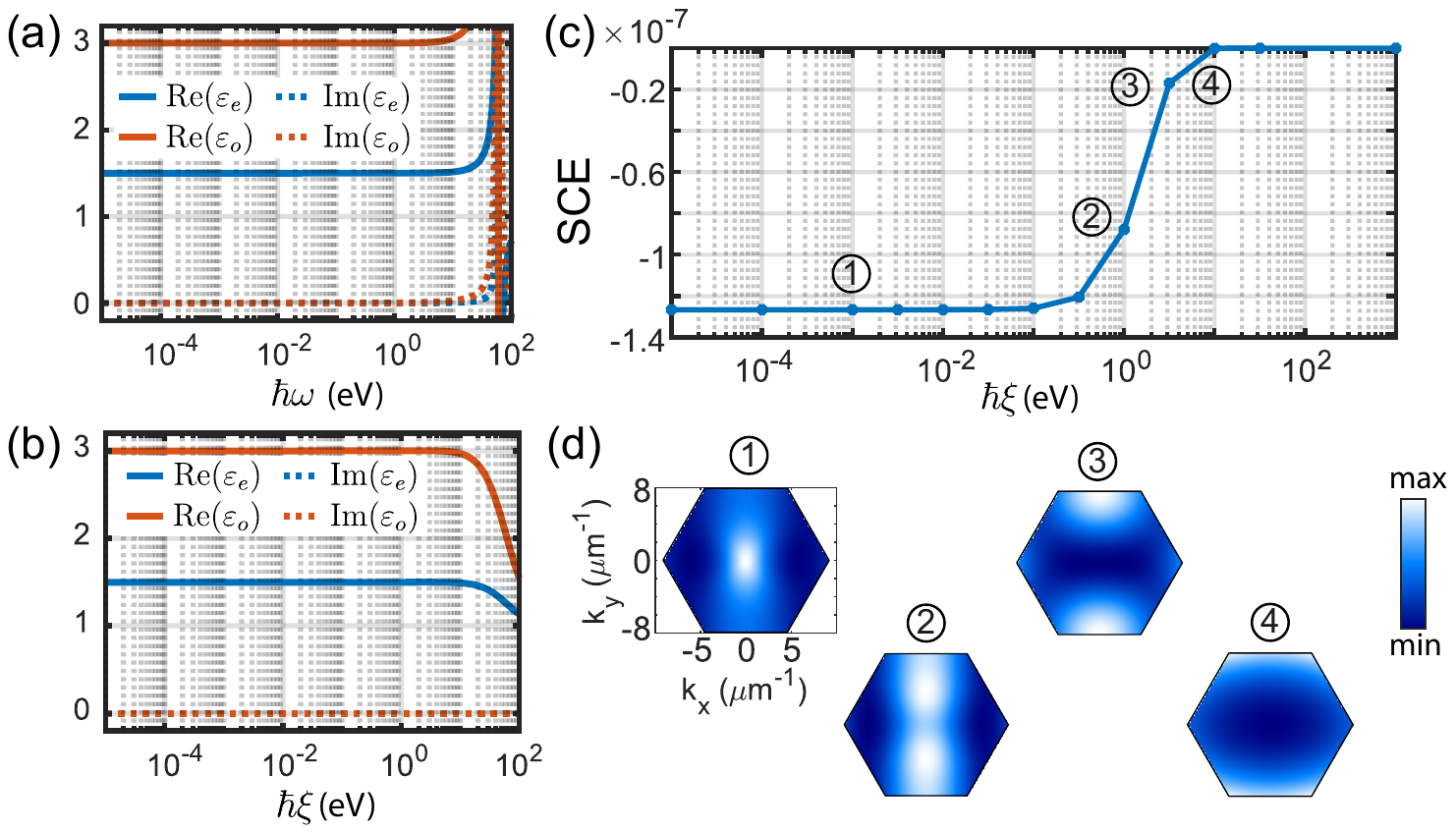}
    \caption{(a)--(b) Dielectric permittivity components $\varepsilon_e$ and $\varepsilon_o$ at real and imaginary frequencies. (c) Spectral Casimir energy (SCE) as a function of imaginary frequency. (d) Spectral modal Casimir energy (SMCE) in the first Brillouin zone for four frequencies. Results for the panels (c) and (d) are calculated for $\alpha = 60^{\circ}$, $\theta = 60^{\circ}$ and $g = 100$~nm.}
    \label{fig_end}
\end{figure*}

The dielectric material that we consider experiences negative anisotropy $\varepsilon_o > \varepsilon_e$, which means that the out-of-axis component dominates over the axial one. We selected these parameters to maximize the anisotropic contrast, thereby strengthening the Casimir interaction. 
The qualitative behavior of the effect remains similar for materials with positive anisotropy.  When analytically continued to the imaginary frequency axis ($\omega \rightarrow i\xi$), both dielectric components $\varepsilon_e(i\xi)$ and $\varepsilon_o(i\xi)$ become real-valued, monotonically decreasing functions with vanishing imaginary parts. This is illustrated in Fig.~\ref{fig_end}(a) for real frequencies and in Fig.~\ref{fig_end}(b) for imaginary frequencies, respectively. To introduce asymmetry into our structure, we rotate the anisotropy axis by angle $\theta$ within the plane of the grating. The rotated dielectric tensor for the non-rotated grating (i.e., $\alpha = 0^{\circ}$, where the grating stripes are aligned along the $y$-axis) is expressed as:
\begin{equation}
    \hat{\varepsilon}^{(i)}(\theta) = \begin{pmatrix}
    \varepsilon_{xx}^{(i)}&\varepsilon_{xy}^{(i)}&0\\ \varepsilon_{yx}^{(i)}&\varepsilon_{yy}^{(i)}&0\\
    0&0&\varepsilon_{zz}^{(i)}
    \end{pmatrix},
    \label{ten2}
\end{equation}
where $i = 1,2$ corresponds to the 1-st or 2-nd grating and:
\begin{align*}
\varepsilon_{xx}^{(i)} &= \varepsilon_e\cos{\theta_i}^2+\varepsilon_o\sin{\theta_i}^2 \\
\varepsilon_{yy}^{(i)} &= \varepsilon_o\cos{\theta_i}^2+\varepsilon_e\sin{\theta_i}^2\\
\varepsilon_{xy}^{(i)} &= \varepsilon_{yx}^{(i)} = (\varepsilon_o-           \varepsilon_e)\sin{\theta_i}\cos{\theta_i} \\
\varepsilon_{zz}^{(i)} &= \varepsilon_o
    \label{ten2}
\end{align*}

\section{Spectral Modal Casimir energy and Spectral Casimir Energy}

The function being integrated over the Brillouin zone in Eq.~2 in main text of the manuscript has a meaning of the Spectral Modal Casimir Energy (SMCE); it depends on $\mathbf{k}$ and $i\xi$. The integration of the SMCE over the $\mathbf{k}$, in turn, represents a Spectral Casimir Energy (SCE); it depends only on $i\xi$.

Fig.~\ref{fig_end}(d) displays the SMCE as a function of the in-plane wavevector within the first Brillouin zone for various frequencies. All functions are smooth and lack singularities, guaranteeing robust numerical integration for a given $i\xi$. Furthermore, Fig.~\ref{fig_end}(c) shows the SCE as a function of imaginary frequency $i\xi$. The SCE remains constant for energies up to $0.1$~eV, after which its absolute value decreases, approaching zero for frequencies above $10$~eV. This demonstrates that the SCE integrand is smooth and localized in the $i\xi$ domain, further ensuring efficient and accurate computation for a given geometry.

Although the integration of the SMCE over the in-plane wavevector at a given frequency is fast and robust, the full of the Casimir energy calculation becomes computationally demanding. This is due to the necessity of integrating the resulting SCE over the imaginary frequency domain, a process that must be repeated for multiple twist angles, anisotropy axis rotation angles, and gap sizes. To manage this high computational cost, we employ the following strategy: exhaustive calculations are performed only for a discrete set of rotation angles, while the results for intermediate angles are then approximated by interpolation with analytical functions, as was shown.
\section{Chiral configuration of identical gratings }

In the main text we analyzed the Casimir interaction for two representative cases: the \textit{symmetric} and the \textit{chiral} configurations. The symmetric case corresponds to stacks of gratings with the anisotropy axis oriented at $\theta = 0^{\circ}$ or $\theta = 90^{\circ}$, where each grating possesses vertical mirror-symmetry planes. In contrast, the chiral case arises when this symmetry is broken, for deviations of the anisotropy axis from the symmetric orientations, $\theta \neq 0^{\circ},90^{\circ}$. In the examples discussed in the main text, the upper and lower gratings were taken with opposite orientations, $\theta_1 = -\theta_2 = \theta$. 

\begin{figure*}[ht]
    \centering
    \includegraphics[width=1\linewidth]{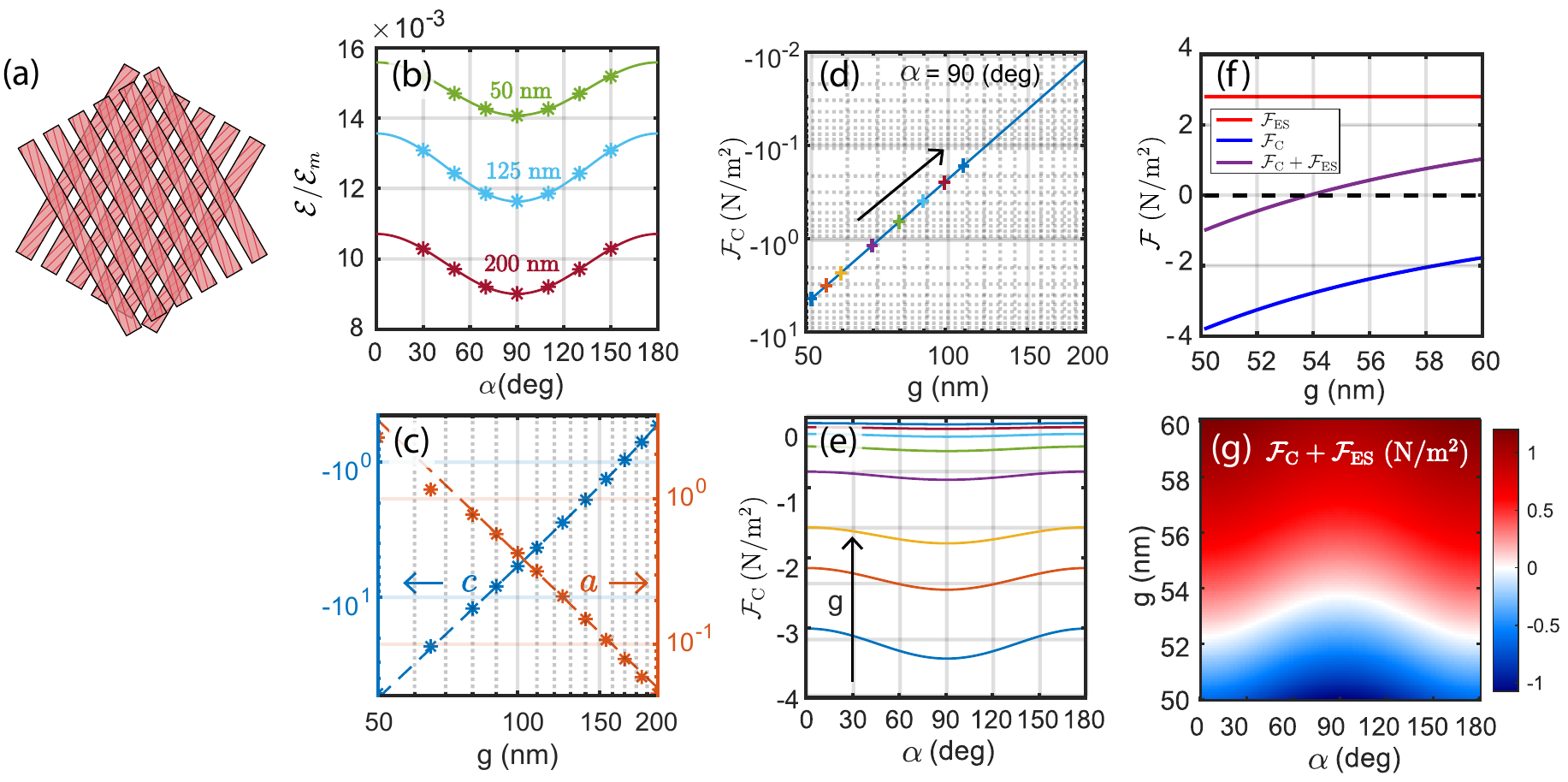}
    \caption{(a)Chiral configurations of twisted stack on to identical grating with in-plane chirality. (b) The twist-angle dependence of the Casimir energy normalized to that for two ideal metallic plates for different gap sizes. (c) The gap-size dependencies of the fit parameters $a(g)$ and $c(g)$.  The Casimir force (d) versus gap $g$ at $\alpha=90^\circ$ and (e) versus $\alpha$ for distances marked in (d).  (f) The Casimir, electrostatic and resultant forces dependence on the distance $g$. (g) The resultant of Casimir force and electrostatic force versus gap $g$ and angle $\alpha$. Results are computed for $\theta = 30^{\circ}$.}
    \label{supp_fig1}
\end{figure*}

Here Fig.~\ref{supp_fig1} we extend the analysis to the case of \textit{identical} gratings with in-plane chirality, $\theta_1 = \theta_2 = \theta$. The Casimir interaction in this configuration exhibits the same qualitative behavior as in the symmetric case: the energy minimum remains stable at $\alpha = 0^{\circ}$, and both the Casimir force and the combined Casimir–electrostatic force show a similar angular dependence.

\section{Temperature dependence of Casimir effect}

\begin{figure*}[h!]
    \centering
    \includegraphics[width=1\textwidth]{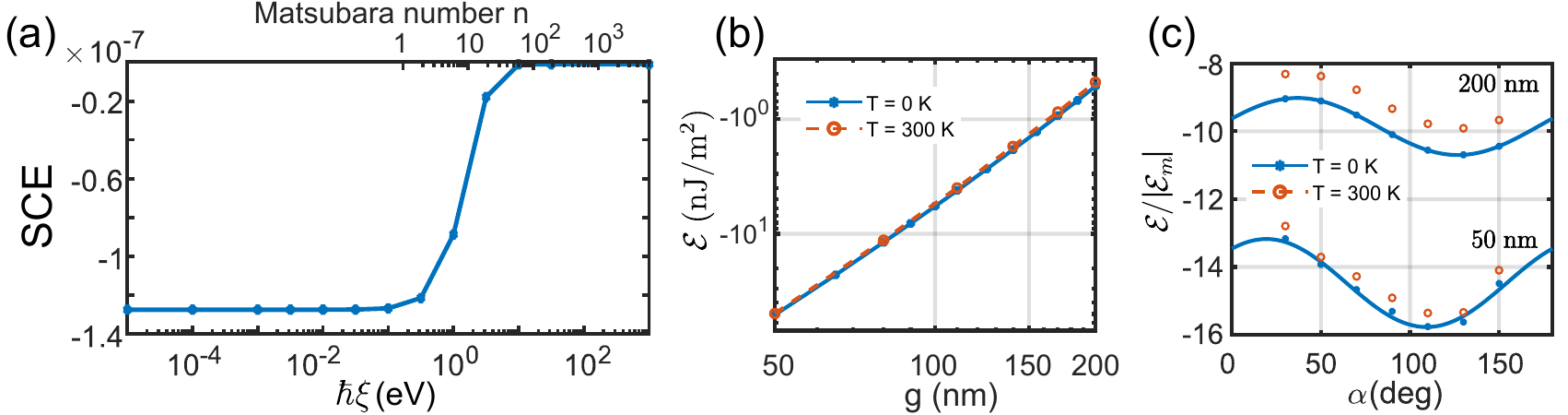}
    \caption{Comparison of the Casimir energy in the zero-temperature limit and at room temperature. 
    (a) Spectral Casimir energy (SCE) as a function of imaginary frequency and Matsubara numbers $n$. 
    (b) Distance dependence of the Casimir energy for $\theta = 30^{\circ}$ and $\alpha = 70^{\circ}$. 
    (c) Dependence of the Casimir energy on the twist angle $\alpha$, normalized to the absolute value of the Casimir energy between two ideal metallic plates, for separations of $g = 50$~nm and $g = 200$~nm.}
    \label{supp_fig2}
\end{figure*}

For nonzero temperatures $T$, thermal fluctuations contribute to the Casimir interaction in addition to quantum fluctuations. In this case, the continuous integral over the imaginary frequency $\xi$ is replaced by a discrete sum over Matsubara frequencies $i\xi_n = i\,2\pi n k_B T / \hbar$ \cite{barash1978moment,guerout2013thermal}, as illustrated in Fig.~\ref{supp_fig2}(a):
\begin{equation*}
    \frac{\hbar}{2\pi}\int_{0}^{\infty} \mathrm{SCE}(\xi)\, d\xi
    \;\rightarrow\;
    k_B T \sum_{n=0}^{\infty'} \mathrm{SCE}(\xi_n).
\end{equation*}
Here, $k_B$ is the Boltzmann constant, and the prime on the summation indicates that the $n=0$ term is taken with a weight of $1/2$.

In this study, we consider an artificial material whose dielectric response is similar to that of common optical glasses and exhibits only a weak temperature dependence. Under the assumption of temperature-independent optical properties, we additionally calculated the Casimir energy at room temperature ($T = 300$~K) and compared it with the zero-temperature results presented in the main manuscript. Figure~\ref{supp_fig2} shows the Casimir energy (b) as a function of the separation distance $g$ for a representative configuration with the anisotropy axis tilted by $\theta = 30^{\circ}$ and a relative rotation angle between the gratings of $\alpha = 70^{\circ}$, and (c) as a function of the relative rotation angle $\alpha$ for separation distances of $g = 50$~nm and $200$~nm.

A comparison between the zero-temperature and room-temperature Casimir energies reveals that (b) the distance-dependent curves nearly coincide and (c) the angular dependence remains unchanged. These results confirm that thermal effects do not introduce significant modifications within the considered parameter range, thereby validating the use of the zero-temperature approximation in our analysis. 


\section{Electrostatic force between charged gratings}

To calculate the electrostatic force we suppose that the dielectric stripes of each grating carry a uniform surface charge density $\sigma$. 

Let a point charge $q_0$ be located at the origin in free space. At a field point $\mathbf r=(x,y,z)$ with $r=|\mathbf r|=\sqrt{x^2+y^2+z^2}$ 
the Coulomb law gives
\begin{equation}
  \mathbf E(\mathbf r)
= \frac{1}{4\pi\varepsilon_0}\frac{q_0 \mathbf{r}}{r^3}
\end{equation}


Further, consider an infinite straight chain of charges lying on the $y$-axis with uniform linear density $\lambda$. Let $\rho=\sqrt{x^2+z^2}$ denote the radial distance from the axis.

The contribution of an element $\lambda dy_0$ located at $(0,y_0,0)$ to the field at point$(x,y,z)$ is
\begin{equation}
dE_\rho = \frac{\lambda}{4\pi\varepsilon_0}
\frac{\rho dy_0}{(\rho^2+(y-y_0)^2)^{3/2}}.
\end{equation}
Integration over $y_0\in(-\infty,\infty)$ yields
\begin{equation}
E_\rho(\rho)=\frac{\lambda}{2\pi\varepsilon_0\,\rho}.
\end{equation}
We consider only the radial component, because due to symmetry, the components along the $y$-axis are compensated. 

we can compare this result to the one obtained with Gauss's law. Choose a cylindrical Gaussian surface of radius $\rho$ and length $L$ 
coaxial with the chain. The enclosed charge is $\lambda L$. 
The flux is $E_\rho(2\pi\rho L)$. 
Thus
\begin{equation}
\varepsilon_0 E_\rho(2\pi\rho L)=\lambda L 
\quad\Rightarrow\quad
E_\rho(\rho)=\frac{\lambda}{2\pi\varepsilon_0\,\rho}.
\end{equation}

Now, let chains (each with density $\lambda$) fill the interval $x_0\in[-w/2,w/2]$ in the plane $z=0$. It is convenient to introduce a surface charge density (charge per unit area) $\sigma$ defined by $\sigma(x_0) = \lambda\times \lambda_x$, where $\lambda_x$ is the number of chains per unit $x_0$. We postulate the $\sigma$.
At a point $(x,0,z)$ the differential contribution from a chain at $x_0$ is
\begin{equation}
dE_x=\frac{\sigma dx_0}{2\pi\varepsilon_0}\,
\frac{x-x_0}{(x-x_0)^2+z^2},\qquad
dE_z=\frac{\lambda dx_0}{2\pi\varepsilon_0}\,
\frac{z}{(x-x_0)^2+z^2}.
\end{equation}
Hence the total field is
\begin{equation}
E_x(x,z)=\frac{\sigma}{2\pi\varepsilon_0}
\int_{-w/2}^{w/2}\frac{x-x'}{(x-x')^2+z^2}\,dx',
\end{equation}
\begin{equation}
E_z(x,z)=\frac{\sigma}{2\pi\varepsilon_0}
\int_{-w/2}^{w/2}\frac{z}{(x-x')^2+z^2}\,dx'.
\end{equation}
Evaluation gives
\begin{multline}
E_x(x,z)=\frac{\sigma}{4\pi\varepsilon_0}\,
\ln\!\frac{(x-w/2)^2+z^2}{(x+w/2)^2+z^2},\\
E_z(x,z)=\frac{\sigma}{2\pi\varepsilon_0}\left(
\arctan\frac{x-w/2}{z}-\arctan\frac{x+w/2}{z}\right).
\end{multline}

\begin{figure*}[ht]
    \centering
    \includegraphics[width=1\linewidth]{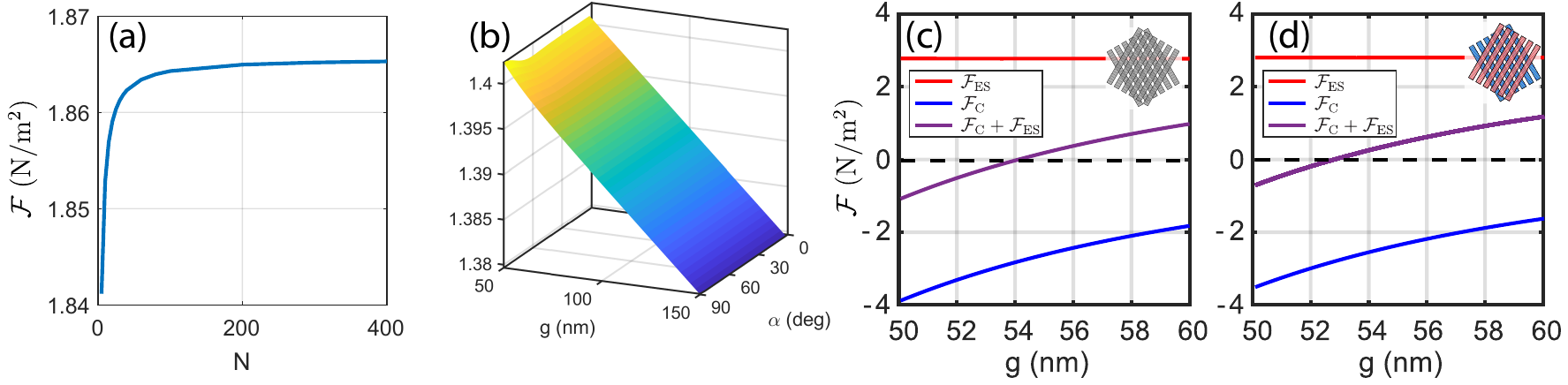}
    \caption{(a) Convergence of the expression Eq.11 according to the number of stripes in sum ($n \in [-\text{N},\text{N}]$). (b) The dependence of the electrostatic force Eq.11 on the twist angle $\alpha$ and distance $g$. (c)--(d) The gap-size dependencies of the Casimir, electrostatic and resultant forces for stacks of symmetric lattices and in-plane chiral lattices. Results are computed for $\theta = 30^{\circ}$ and charge density $\sigma = 10^{-5}$ C/m$^2$.}
    \label{supp_fig3}
\end{figure*}

Finally, consider an infinite array of parallel strips, each of width $w$, aligned with the $y$-axis and centered at positions $X_n=n p$, $n\in\mathbb Z$, 
with period $p$ along the $x$-direction. Each strip carries a surface charge with density $\sigma$.
At an observation point $(x,z)$ with $z=g$, the $z$-component of the electric 
field is obtained by superposition:
\begin{equation}
E_z(x,z)=\frac{\sigma g}{2\pi\varepsilon_0}
\sum_{n=-\infty}^{\infty}\int_{-w/2}^{w/2}
\frac{dx_0}{(x-x_0-X_x)^2+g^2}.
\end{equation}
based on the previous result Eq.(8):
\begin{multline}
E_z(x,z)=\frac{\sigma}{2\pi\varepsilon_0}\sum_{n=-\infty}^{\infty}\left( \arctan\frac{x-w/2-X_n}{g}- \right.
\\
\left.-\arctan\frac{x+w/2-X_n}{g} \right).    
\end{multline}
This sum converges rapidly with $n$ (see Fig.~\ref{supp_fig3} (a)), so a finite number of terms can be retained without loss of generality. 


To compute the force on a stripe located in the plane $z = g$, one multiplies the electric field by the charge distribution on the second stripe. Since the integral over $y$ diverges, it is convenient to evaluate the force per unit length of the stripe (along $y$) or per unit area.  

First, let us consider a stripe centered at $x = 0$ and oriented parallel to the stripes in the plane $z = 0$. The force per unit area can then be calculated using the following expression:  
\begin{multline}
    F_z(x,z)=\frac{\sigma^2}{2\pi\varepsilon_0 p}\sum_{n=-\infty}^{\infty}\int_{-w/2}^{w/2}\left(
\arctan\frac{x-w/2-X_n}{g}-\right.
\\
\left.-\arctan\frac{x+w/2-X_n}{g} \right) dx.
\end{multline}

This integral is evaluated numerically using MATLAB. If the center of the stripe is shifted by a distance $s$ from $x = 0$, the variable $x$ in the expression should be replaced by $x+s$. The force on each stripe in the upper plane is the same.

If the upper stripe is rotated by an angle $\alpha$ with respect to the stripes in the lower plane, an elementary charge element $\sigma\,dx\,dy$ is located, in the original coordinates, at the point $(x\cos\alpha - y\sin\alpha,\, x\sin\alpha + y\cos\alpha,\, g)$. Substituting these coordinates into Eq.~(10) and integrating over the area overlapping with a single period of the lower lattice, we obtain the limits $x \in [-w/2,\, w/2]$ and $y \in [-p/(2\sin\alpha), p/(2\sin\alpha)]$. The result is then normalized by the area $S = p^{2}/\sin\alpha$. 

\begin{multline}
F_z(x,z)=\frac{\sigma^2\sin(\alpha)}{2\pi\varepsilon_0 p^2}\sum_{n=-\infty}^{\infty}\int_{-w/2}^{w/2}\int_{-p/(2\sin\alpha)}^{p/(2\sin\alpha)}\\\left(
\arctan\frac{x'-w/2-X_n}{g}-\arctan\frac{x'+w/2-X_n}{g}\right)dxdy
\end{multline}
where $x' = x\cos\alpha - y\sin\alpha$.
As shown in Fig.~\ref{supp_fig3}(b), the averaged force is nearly independent of the rotation angle $\alpha$. Therefore, in the main text the calculations were carried out for a case with $\alpha = 90^{\circ}$. Figures~\ref{supp_fig3}(c)--(d) present the Casimir force, the electrostatic force, and their resultant. The value of charge density in presented calculations is $\sigma = 10^{-5}$ C/m$^2$. The electrostatic contribution ($\mathcal{F}_\text{ES} = F_z$, see Eq.~12) decays much more slowly than the Casimir force, leading to an unstable equilibrium at the distance where the total force vanishes.

\end{document}